%% file: final.tex
\RequirePackage{fix-cm}
\documentclass[smallextended]{svjour3} 
\smartqed 
\include{PSIpackages}
\include{PSImacros}

\usepackage{graphicx}
\begin{document}

\title{How to Measure the Quantum Measure 
}

\subtitle{\bigskip\centerline{\it\/In memory of David Ritz Finkelstein}}


\author{\'Alvaro Mozota Frauca \and
Rafael Dolnick Sorkin 
}



\institute{\'Alvaro Mozota Frauca \at
Perimeter Institute for Theoretical Physics, Waterloo, ON N2L 2Y5, Canada \\
\and
Rafael Dolnick Sorkin \at
Perimeter Institute for Theoretical Physics, Waterloo, ON N2L 2Y5, Canada \\
Department of Physics, Syracuse University, Syracuse, NY 13244-1130, U.S.A.
}

\date{Received: date / Accepted: date}

\maketitle

\begin{abstract}
The histories-based framework of Quantum Measure Theory assigns a
generalized probability or {\it\/measure\/} $\mu(E)$ to every (suitably
regular) set $E$ of histories. Even though $\mu(E)$ cannot in general be
interpreted as the expectation value of a selfadjoint operator (or
POVM), we describe an arrangement which makes it possible to determine
$\mu(E)$ experimentally for any desired $E$.
Taking, for simplicity, the system in question to be a particle passing
through a series of Stern-Gerlach devices or beam-splitters, we show how
to couple a set of ancillas to it, and then to perform on them a suitable
unitary transformation followed by a final measurement, such that the
probability of a final outcome of ``yes'' is related to $\mu(E)$ by a known
factor of proportionality.
Finally, we discuss in what sense a positive outcome of the final
measurement should count as a minimally disturbing verification that the
microscopic event $E$ actually happened.
\end{abstract}

\section{Introduction}
Ever since quantum theory was first put into the form of a complete
mathematical scheme, there have been innumerable attempts to explain it
and to understand what it is trying to tell us about the world.
Depending on what version of quantum mechanics one follows, and how one
interprets it, one needs to abandon one or another of the classical
ideas we are comfortable with, such as causality, locality, or
homomorphic logic \cite{quant-ph/0703276}. Perhaps the central question
that everyone faces is the so-called measurement problem, the fact that
the theory appears to assert that when someone measures a system, its wave
function ``collapses''. No longer a superposition, it now corresponds
to a definite value for the physical property that has been measured and
which the system has somehow acquired.

Different interpretations explain this ``collapse'' differently, mostly
by trying to explain it away. On the on the one hand, one could think
that it is simply the way we update our description when we obtain new
knowledge, that the physical properties characterising a system are
always well defined, but we only learn about them when measuring. Such
interpretations include hidden-variable theories and Bohmian
Mechanics \cite{PhysRev.85.166, PhysRev.85.180}. On the other hand, one
could think that all the possible outcomes of a measurement actually
occur, with the universe branching into multiple realities each time a
measurement is performed. This is the many-worlds
idea \cite{RBrown2005}. Or one could argue that the collapse is an
illusion stemming from the decoherence that takes place when a system
interacts with a measuring device or with an environment.

In this paper we will work in the framework of Quantum Measure Theory
\cite{gr-qc/9401003, gr-qc/0403085,quant-ph/0605008, Gudder2011, gr-qc/9507057},
which generalizes the mathematical concept of a measure-space so as to
allow for quantal interference. When refashioned in this language,
Quantum Mechanics appears as a generalized probability theory of a type
inspired by the path integral. Instead of a wave function, Quantum
Measure Theory works with the histories of the system, assigning a real
number, the measure $\mu$, to every set of histories. In some special
measurement-situations, $\mu$ gives the probability of the outcome of an
experiment one could perform, but in general one can not identify it
with any observable probability.

The idea behind this reformulation is to arrive at an understanding
which not only provides probabilities for certain types of laboratory
events, but which goes further by offering a framework within which one
can speak about the microworld directly, without needing to presuppose
concepts such as experiment, observer or measurement. To that end,
several schemes have been proposed in which reality is described by a
certain mathematical combination of individual histories (individual
particle trajectories for example) called a ``coevent''. Since multiple
histories enter into this description of reality, one could be tempted
to fit Quantum Measure Theory into the many-worlds interpretation. Or
because it works with definite trajectories,
one could also be tempted to fit it into Bohmian Mechanics.
In fact, however, Quantum Measure Theory doesn't fit into any of these
interpretations, and offers a distinctive vantage point from which one
can view the measurement problem.

Quantum Measure Theory is also intended to be the right dynamical
framework for Quantum Gravity. For the Causal Set
approach \cite{PhysRevLett.59.521} in particular, it provides a
dynamical law which can describe the growth of the causal set, of the
universe, without succumbing to the limitations which the
Schr{\"o}dinger equation encounters when a continuous, background time
is unavailable. The ability to do without a fundamental notion of
measurement or external agent is likewise important to a theory like
quantum cosmology, whose field of application is one where no
recognizable ``observer'' could exist
\cite{1103.6272, gr-qc/9904062, gr-qc/0508109,gr-qc/0309009}.

A question that arises naturally in connection with Quantum Measure
Theory is whether the measure $\mu$ has any experimental significance
outside the special context in which it can be interpreted as the
Born-rule probability of a particular instrument-event. To the extent
that it does, this will enhance its status as an independent way to
formulate quantum mechanics, and it will also suggest practical
experiments which would test quantum predictions about events of a
different type than one usually deals with.

The goal of this paper is to provide a positive answer to the question
just raised.
We will present schematically an experimental setup that will reveal the
measure of any given set of histories (any given event), including
events extended arbitrarily in time. As we have said, not every event
$E$ can be made to correspond with a projection operator (or member of a
POVM) whose expectation value would be $\mu(E)$. To compensate for this
we will need to couple the system to suitable ancillas and then to
perform suitable transformations on them followed by a final projective
measurement. But (perhaps surprisingly) we will not require anything
more exotic.
The procedure we will describe
may be thought of
as a way to filter which
trajectories a particle can have travelled,
based on a generalization of the Quantum Eraser \cite{PhysRevA.25.2208,
PhysRevLett.75.3034, Aharonov875}.
Whether in Quantum Mechanics we can speak about particle trajectories as
we are accustomed to do classically is not something that everyone
agrees on \cite{PhysRevLett.111.240402, PhysRevA.87.052104}, but our
results will illustrate how one can do so consistently in Quantum Measure
Theory.

The plan of this paper is the following. First we introduce Quantum
Measure Theory ($\S$ \ref{sect2}), then we introduce the system we will
study ($\S$\ref{sect3}), and then we explain how to couple our ancillas
to the system ($\S$\ref{sect4}) and how to process them so as to obtain
the measure we are looking for ($\S$\ref{sect5} and $\S$\ref{sect6}).
Finally we will suggest how to interpret our results ($\S$ \ref{sect7})
and conclude with some summary remarks and possible extensions of our
work ($\S$\ref{sect8}).

This paper is dedicated in memory of David Finkelstein, whose thought
continues to guide fundamental physics more than most workers probably
appreciate. For RDS especially, David was a mentor and inspiration from
graduate school days onward, and from Manhattan to Athens to Atlanta.
We like to think that David, who once wrote
``I attach observables to histories, not instants'',
would have been pleased to see how this declaration of his
might be put into practice.

\section{Quantum Measure Theory \label{sect2}}
A measure on a space $\Omega$ is a way to assign a number to each
suitable subset of $\Omega$. An example of a classical measure is the
probability measure on a sample space, or the Lebesgue measure on a
Euclidean space which, depending on the dimension $n$, gives
to each measurable subset of $\R^n$.
its conventional length, area, volume or hyper-volume in Euclidean geometry.

In the classical case, a measure space is defined formally by the triple
formed by a set $\Omega$, a set-algebra $\mathcal{A}$ over $\Omega$ and
a function $\mu:\mathcal{A} \longmapsto \R^+$. A set-algebra over a set
$\Omega$ is a set of subsets of $\Omega$, including the empty set, and
closed under complementation, union, and intersection. (In the
classical case, one usually requires also closure under infinite
sequences of intersections or of unions, making $\mathcal{A}$ a
$\sigma$-algebra.) The function $\mu$ is called the measure.

Quantum Mechanics can be understood as a generalized measure theory on
the space $\Omega$ of possible {\it\/histories\/} of some physical system. It
assigns a non-negative real number to every {\it\/event\/}, an event
being a subset of $\Omega$, in other words a set of histories. The
``quantum measure'' $\mu$ that does this
cannot be an ordinary probability measure
because there is interference, in consequence of which $\mu$ is neither
additive nor bounded above by unity. It is a ``generalized measure''
for which the measure of an event is not simply the sum of the
probabilities of the histories that compose it.
Instead, the measure of an event is given (in an extension of the Born
rule to general events) by the sum of the squares of certain sums of the
complex amplitudes of the histories which comprise the event.

As just stated, $\Omega$ is the history-space of the physical system in
question. By history we mean a complete classical description of
the physical reality of our system, for example a particle's history
would be its trajectory or worldline, while a field's history would be
its configuration in spacetime.
Knowing the measures of sets of histories (knowing $\mu (A)$ $\forall A
\in\mathcal{A}$) allows you to make predictions about the system in a
similar way to how, in the usual formulation of Quantum Mechanics,
knowing the wavefunction allows you to make predictions. Moreover,
there exist quantal measures that yield theories more general than
Quantum Mechanics, for example non-unitary theories. \cite{gr-qc/0403085}

As we have said, the feature that distinguishes a quantum theory from a
classical theory is interference. This means that the measure will enjoy
different formal properties than classically.
We can define the following set-functions for
any generalized measure theory over a sample space $\Omega$:
\begin{equation}
I_1(A)=\mu (A)
\end{equation}
\begin{equation}
I_2(A,B)=\mu (A\cup B)-\mu (A)-\mu (B)
\end{equation}
\begin{equation}
I_3(A,B,C)=\mu (A\cup B\cup C)-\mu (A\cup B)-\mu (B\cup C)-\mu (A\cup C)+\mu (A)+\mu (B)+\mu(C)
\end{equation}
and so on, where $A, B, C, etc.$ are disjoint subsets of $\Omega$.

These functions allow us to distinguish between different types of theories. We will say
that a theory is of level $k$ if it satisfies $I_{k+1}=0$. One can show
that this condition implies also $I_m=0$ for every $m$ bigger than $k+1$. A
classical theory is one of level 1, which is equivalent to saying that there is no
interference: $\mu (A\cup B)=\mu (A)+\mu (B)$.
A {\it\/quantum measure theory\/} is a
theory of level 2, i.e. a theory with second order but no higher order
interference.
An example
is ordinary quantum mechanics, but
it is not the only class of
theories
in this category.

Beyond level 2, several researchers have been investigating the
possibility of theories residing at level 3 or higher \cite{H1,H2,H3,H4,H5,H6,H7,H8},
but for the moment there has not been any evidence of higher-order
interference from the experiments that have looked for it. See for
example the three-slit experiments that have put increasingly stringent
bounds on third order interference.
Such theories are,
in any case,
outside the scope of this paper.

Any normalized quantum measure can be built by using a decoherence functional
$D: \mathcal{A} \times \mathcal{A} \longmapsto \C$ on pairs of subsets of
$\Omega$ which satisfies:
\begin{equation}
\label{herm}
D(A;B)=\overline{D(B;A)} \qquad \forall A, B
\end{equation}
\begin{equation}
D(A\cup B;C)=D(A;C)+D(B;C) \qquad \forall A, B, C \quad \text{with} \quad A,B \quad \text{disjoint}
\end{equation}
\begin{equation}
D(A;A)\geq 0 \qquad \forall A
\end{equation}
\begin{equation}
\label{norm}
D(\Omega ;\Omega )= 1
\end{equation}
The quantum measure in terms of the decoherence functional is:
\begin{equation}
\mu (A) =D(A;A)
\end{equation}
One can check that any measure defined this way is a level 2 measure.

Using these ideas,
let us see how,
via the path-integral,
ordinary Quantum Mechanics can be understood as a level-2 measure theory.
In ordinary quantum mechanics
the probability of an experimental outcome (probability density for
continuous outcomes), let's say a particle being at a position $x_0$ at a time
$t_0$,
is supposed to be given
by the square of the amplitude associated with that event. For a particle this
amplitude is given by the wave function, or equivalently by a path integral over
all possible histories ending with the particle at $x_0$ at $t_0$.
\begin{equation}
\label{psi}
\Psi(x_0,t_0)=\int^{x(t_0)=x_0} \mathcal{D}[x] e^{iS[x]/\hbar}
\end{equation}
\begin{equation}
\label{psisq}
p(x_0,t_0)=|\Psi(x_0,t_0)|^2=\int^{x(t_0)=x_0}\int^{y(t_0)=x_0} \mathcal{D}[x] \mathcal{D}[y] e^{iS[x]/\hbar}e^{-iS[y]/\hbar}
\end{equation}
In this expression, which implicitly contains the Born rule,
the amplitude of each individual history is given by the exponential of
$iS[x]/\hbar$, $S[x]$ being the action evaluated along the trajectory.
One can verify that a wavefunction defined via (\ref{psi})
evolves unitarily, obeying the
Schr\"odinger equation with the Hamiltonian associated with the
action $S$. Thus, the wavefunction formalism is
in a sense
contained in
the path integral formalism.

Now, we want to show that our double path integral for the
probability-density is equivalent to a level 2 measure. For doing so we
define the following decoherence functional for a pair of histories:
\begin{equation}
D(x;y)=e^{iS[x]/\hbar}e^{-iS[y]/\hbar}\delta_{x(t_0),y(t_0)}
\end{equation}
Here $x$ and $y$ denote two histories and we have made explicit
the delta function of
the final positions that is implicit in (\ref{psisq}).
(This condition that only histories that end at the same point can
interfere, might seem to give a special status to the ``collapse time''
$t_0$. That $\mu$ nevertheless be independent of $t_0$, implies a
consistency condition which holds automatically, thanks to unitarity.)
The decoherence functional evaluated
on general sets
can be derived by using the formal properties
\eqref{herm}-\eqref{norm}:
\begin{equation}
D(X;Y)=\int _{x \in X}\int _{y \in Y}\mathcal{D}[x]\mathcal{D}[y] e^{iS[x]/\hbar}e^{-iS[y]/\hbar}\delta_{x(t_0),y(t_0)}
\end{equation}
(Instead of a sum over the trajectories contained in the sets $X$ and
$Y$, we have an integral, because we are working with continuous
variables.)
Now we can compute the measure of the set $X(x_0,t_0)$, which we
define as the set of all possible histories ending at $(x_0, t_0)$:
\begin{equation}
\begin{split}
\mu(X(x_0,t_0))
&= D(X(x_0,t_0);X(x_0,t_0)) dx_0
\\&= \int^{x(t_0)=x_0}\int^{y(t_0)=x_0}\mathcal{D}[x]\mathcal{D}[y] e^{iS[x]/\hbar}e^{-iS[y]/\hbar}
\end{split}
\end{equation}
We have thus recovered
from the decoherence functional
the same probability density
that one computes using ordinary quantum
mechanics. In this manner, one
can understand Quantum Mechanics as a level 2 measure theory.

In the formulas just above, we had continuous integrals,
but for discrete systems, we will have sums, in which case
the decoherence functional will take the simpler form
\begin{equation}
D(X;Y)=\sum _{x \in X,y \in Y} A(x)\bar A(y)\delta_{x(t_0),y(t_0)}
\end{equation}
where $A(x)$ is the amplitude of the history $x$,
and where $\delta_{x(t_0),y(t_0)}$ now denotes a Kronecker delta.
This will be the applicable form in the remainder of this paper.

We have just seen how
the probability of one particular experimental observable
(the position of a particle at a specified time)
can be understood as the measure of a certain set of histories,
but this is only a start.
There are many other sets of histories (many other events) that do not correspond to
any particular time or any obvious observable of ordinary quantum mechanics.
How to interpret
the measures of such sets is not
evident.
(Recall that a quantum measure $\mu$ can take values bigger than one and
cannot be construed as a probability measure.)
Just in the case of
an event
having a measure 0,
we can say that
the event
does not happen; we will say it is ``precluded''.
But how should we interpret the measure when it does not
vanish? Can its value be made the object of an experimental test?

\section{Our experiment \label{sect3}}
Can we design an experimental setup that
will allow us to ``measure the measure'' of
of any desired event
of a given system. Let us try to find such a setup for the kind of
idealized system one encounters in quantum-information theory and
quantum optics.

Our system will be a particle
that passes
through
a succession of
similar
devices (say Stern-Gerlach analyzers)
which split the beam
into two different trajectories, depending on the eigenvalue of the
observable being ``filtered'',
and
such that the beams are reunited
before the next device
so that they can interfere with each other.
In this setting a history will simply be one of the possible paths the
particle can follow.
If the particle carries spin-$1/2$
(is a two-level quantum system),
then the beam in which it emerges
from a given analyzer can be labelled by the corresponding eigenvalue, letting
us represent a history by a sequence of eigenvalues which
we will sometimes call a ``chain''.

One could also think of each encounter with an analyser as a kind of
measurement, but if one wanted to use that language, a term like ``fake
measurement'' or ``pre-measurement'' would be more appropriate, unless
one inserted a detector into one of the beams to ``collapse the
wavefunction'' and provide irreversible macroscopic information about
which path the system had travelled.

\begin{figure}
\centering
\includegraphics[scale=1]{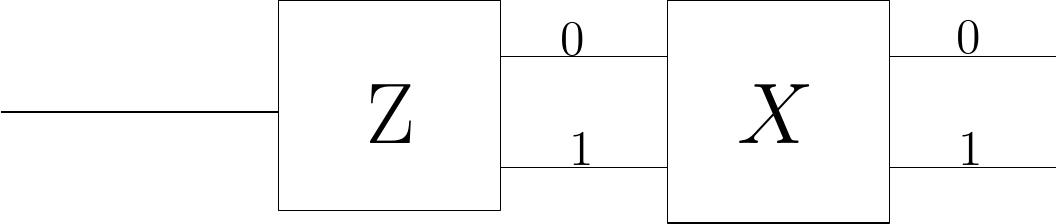}
\caption{\label{box}A simple setup in which we have two Stern-Gerlachs
oriented in the Z and in the X direction. We label with 0 and 1 the
two beams in which the particle can emerge after encountering an analyzer.}
\end{figure}

\begin{figure}
\includegraphics[scale=0.65]{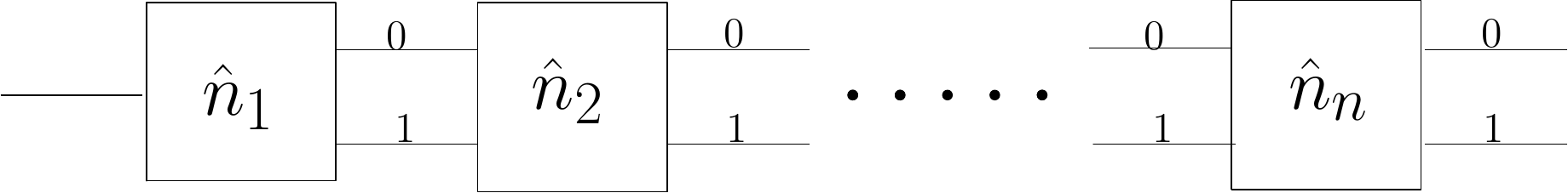}
\caption{\label{boxes}A more general setup with $n$
Stern-Gerlachs, each oriented in its own direction $\hat n_i$. We label
the beams with 0 and 1, as before.}
\end{figure}

\begin{figure}
\centering
\includegraphics[scale=1]{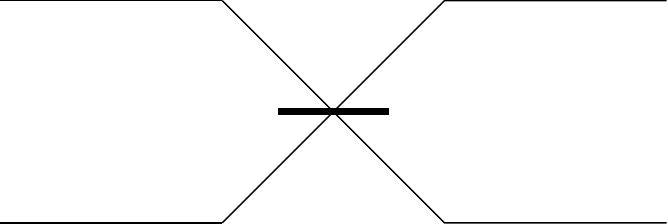}
\caption{\label{optics}A setup in Quantum Optics that is equivalent to the
setup of figure \ref{box}. The thin lines represent paths that
a photon could follow, and the thick line represents a beamsplitter.}
\end{figure}

An example of this kind of setup is a series of Stern-Gerlach apparatuses
oriented in different directions and a spin 1/2 particle travelling through this
series of apparatuses, as we can see in figures \ref{box} and
\ref{boxes}. Another example is an optical circuit like the one shown in figure
\ref{optics}. In this kind of circuit, the
ket $\ket 0$ corresponds to a
photon travelling in the upper branch and $\ket 1$ to a photon travelling in the
lower branch.
In relation with the previous example,
the beam splitter serves the dual purpose of reuniting the two beams and
then splitting them again according to a different eigenbasis.
For reflectivity 1/2,
the setup
is equivalent to the one in figure \ref{box},
since if we identify the ingoing beams (before the beamsplitter) with
eigenstates in the Z-basis,
the outgoing beams will correspond to eigenstates in the X basis.
For splitting according to another basis than X,
we would have to design more
complicated combinations of optical devices
(beamsplitters and phase shifters, mainly).

\begin{figure}
\centering
\includegraphics[scale=0.65]{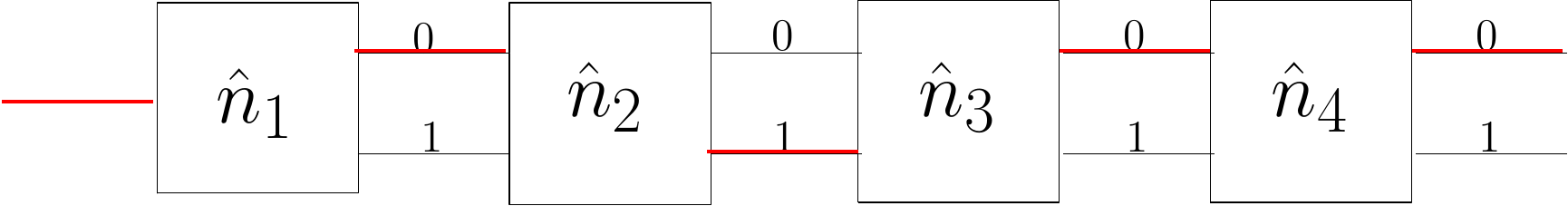}
\caption{\label{Path}In red, an example of a possible path followed by a
particle for a length 4 path. Our notation represents this history as
$\gamma=(0,1,0,0)$}
\end{figure}

For this kind of system, the history formulation is simple.
We can represent a history
$\gamma$ as a chain of $n$ bits for $n$ analyzers,
indicating the corresponding particle-path. We will assign 0 to the
upper beam and 1 to
the lower beam,
and we will write a history as
\begin{equation}
\gamma=(\gamma_1,\gamma_2,\gamma_3...\gamma_n)
\end{equation}
where the $\gamma_i$ are either $0$ or $1$.
An example of how this notation works can be seen in figure \ref{Path}.
From now on we will use the terms ``history'', ``path'', and ``chain'' interchangeably.

For each path, the ordinary quantum mechanical apparatus of state spaces
and projectors gives us an amplitude as follows.
Corresponding to the $i^{th}$ device or ``filter'' is an operator given by $\hat n_i$
(in the Stern-Gerlach case, the direction in which we orient the magnetic field),
and we project the state-vector according to the selected eigenvalue $\gamma_i$:
\begin{equation}
\ket{\Psi_i}=\ket{\hat n_i,\gamma_i}\braket{\hat n_i,\gamma_i}{\Psi_{i-1}}
\end{equation}
where $\ket{\Psi_i}$ and $\ket{\Psi_{i-1}}$ represent the state-vectors after
and before the device, respectively,
and where $\ket{\hat n_i,\gamma_i}$ is the
state-vector with eigenvalue $\gamma_i$ in the $\hat n_i$ direction.
We can expand this result to a chain of length $n$,
$\gamma=(\gamma_1,\gamma_2,\gamma_3...\gamma_n)$ as follows:
\begin{align}
\ket{\Psi_{final}}
=\ket{\hat n_n,\gamma_n}\braket{\hat n_n,\gamma_n}{\hat n_{n-1},\gamma_{n-1}}
\braket{\hat n_{n-1},\gamma_{n-1}}{\hat n_{n-2},\gamma_{n-2}}...\\
...\braket{\hat n_2,\gamma_2}{\hat n_1,\gamma_1}\braket{\hat n_1,\gamma_1}{\Psi_{initial}}
\end{align}
where $\ket{\Psi_{initial}},\ket{\Psi_{final}}$ are the initial and final
state-vectors.
From this, we can read off the amplitude of the chain $\gamma$ as
\begin{equation}
A(\gamma)=\prod\limits_{i=1}^{n} \braket{\hat n_i,\gamma_i}{\hat n_{i-1},\gamma_{i-1}}
\end{equation}
where by $\ket{\hat n_0,\gamma_0}$ we mean the initial wave function $\Psi_{initial}$.

Now that we know the amplitude of a single history, we can compute the
decoherence functional for any two sets of histories,
$X=\{\gamma^{x_1},\gamma^{x_2},\gamma^{x_3},...\}$ and
$Y=\{\gamma^{y_1},\gamma^{y_2},\gamma^{y_3},...\}$ :
\begin{equation}
D(X,Y) =
\sum _{\gamma^{x_k} \in X, \gamma^{y_{k'}} \in Y}
\prod\limits_{i,j=1}^{n}
\braket{\hat n_i,\gamma^{x_k}_i}{\hat n_{i-1},\gamma^{x_k}_{i-1}}
\braket{\hat n_{j-1},\gamma^{y_{k'}}_{j-1}}{\hat n_{j},\gamma^{y_{k'}}_{j}} \delta_{\gamma^{x_k}_n,\gamma^{y_{k'}}_n}
\end{equation}
Setting $X=Y$, we obtain the measure of the event $X$:
\begin{equation}
\label{mu(X)}
\mu(X) =
\sum _{\gamma^{k},\gamma^{k'} \in X}
\prod\limits_{i,j=1}^{n}
\braket{\hat n_i,\gamma^{k}_i}{\hat n_{i-1},\gamma^{k}_{i-1}}
\braket{\hat n_{j-1},\gamma^{k'}_{j-1}}{\hat n_{j},\gamma^{k'}_{j}} \delta_{\gamma^{k}_n,\gamma^{k'}_n}
\end{equation}
We see that $\mu(X)$ is a function of the initial wave function,
of the histories contained in the set $X$,
and of the $n$ settings $n_i$.

\section{Coupling ancillas \label{sect4}}
Now that we have defined the measure of every possible event or set of
histories, we will try to devise a procedure that will
let us determine these measures experimentally.
For that purpose, we will couple in a series of ancillas as follows.
Each ancilla will be prepared in a initial state, $\ket{r}$ where $r$ stands for ready.
At each step the corresponding ancilla will detect which beam the
particle occupies at that point:
\begin{equation}
\ket{0_s}\ket{r}\rightarrow\ket{0_s}\ket{0}
\end{equation}
\begin{equation}
\ket{1_s}\ket{r}\rightarrow\ket{1_s}\ket{1}
\end{equation}
From now on we will distinguish a particle state with the subscript $s$,
leaving the ancilla-states without subscripts. The ready state could be
a third state orthogonal to both $0$ and $1$ (such multilevel ancillas
could be useful if we wanted to couple system to ancilla weakly, as in a
``weak measurement''), but for our purposes, it suffices to make do with
a two level ancilla, with the $0$ state, for example, serving as ready
state. Such a coupling corresponds to a CNOT gate.
For a general superposition, $\ket{\Psi_s}=\alpha \ket{0_s} +\beta \ket{1_s}$,
the ancilla acts as follows:
\begin{equation}
(\alpha \ket{0_s}+\beta \ket{1_s})\ket{r}\rightarrow \alpha\ket{0_s}\ket{0}+\beta\ket{1_s}\ket{1}
\end{equation}

Now suppose we were to measure (``strongly'') the ancilla in the basis, ($\ket{0}, \ket{1}$).
Evidently, this would be equivalent to measuring the particle in the same basis, inasmuch as
the outcome-probabilities would be the same and the particle-state would ``collapse'' in
both cases to the eigenstate associated with the eigenvalue obtained.

If instead we were to measure the ancilla in the basis ($\ket{+}, \ket{-}$) something
different would happen:
\begin{equation}
\ket{+}=\frac{1}{\sqrt{2}}(\ket{0}+\ket{1}) \qquad \ket{-}=\frac{1}{\sqrt{2}}(\ket{0}-\ket{1})
\end{equation}
\begin{equation}
\alpha\ket{0_s}\ket{0}+\beta\ket{1_s}\ket{1} =
\frac{1}{\sqrt{2}} \big[\alpha\ket{0_s}+\beta\ket{1_s}\big]\ket{+} +
\frac{1}{\sqrt{2}} \big[\alpha\ket{0_s}-\beta\ket{1_s}\big]\ket{-}
\end{equation}
\begin{equation}
\text{outcome} \quad \ket+ \quad \rightarrow \quad\quad P(+)=\frac{1}{2}, \quad \ket{\Psi_s}=\alpha \ket{0_s}+\beta \ket{1_s}
\label{o27}
\end{equation}
\begin{equation}
\text{outcome} \quad \ket- \quad \rightarrow \quad\quad P(-)=\frac{1}{2}, \quad \ket{\Psi_s}=\alpha \ket{0_s}-\beta \ket{1_s}
\label{o28}
\end{equation}
As we see from (\ref{o27}) and (\ref{o28}),
the probabilities to obtain $\ket{+}$ or $\ket{-}$ would both be $1/2$,
and we would not learn anything about which path the particle had taken.
Furthermore, after the outcome $\ket{+}$, the wave-function of the system
would have reverted to what it had been before its coupling to the ancilla:
``the system would not have been disturbed'' (``quantum eraser effect'').
On the other hand, after the outcome $\ket{-}$, a phase would have been introduced
into $\ket{\Psi_s}$
in what turns out to be an unhelpful way.

Later we will generalize this result to show that looking for, and finding, a particular
superposition (in this case $\ket{+}$) causes the ancillas `forget' some information,
leaving $\ket{\Psi_s}$ in a ``minimally disturbed'' state.
Notice that this `erasure' is probabilistic; it only succeeds if
we obtain a particular outcome upon measuring the ancillas.

\subsection{Final wavefunction of particle + ancillas}
Now that we have designed the ancilla-system coupling, let's compute the
final wavefunction of the combined system.
Suppose we start with $\ket{\Psi_s}$, and the
$n^{th}$ analyzer is set in the $\hat{n_i}$ direction.
Let's see what happens when we
couple in the first two ancillas:
\begin{align*}
&\ket{\Psi_s}\ket{r}\ket{r} \rightarrow \\
& (\braket{\hat n_1,0_s}{\Psi_s}\ket{\hat n_1,0_s}\ket{0}+\braket{\hat n_1,1_s}{\Psi_s}\ket{\hat n_1,1_s}\ket{1})\ket{r}
\rightarrow \\
& \braket{\hat n_1,0_s}{\Psi_s}\braket{\hat n_2,0_s}{\hat n_1,0_s}\ket{\hat n_2,0_s}\ket{0}\ket{0}
+\braket{\hat n_1,0_s}{\Psi_s}\braket{\hat n_2,1_s}{\hat n_1,0_s}\ket{\hat n_2,1_s}\ket{0}\ket{1}\\
&+\braket{\hat n_1,1_s}{\Psi_s}\braket{\hat n_2,0_s}{\hat n_1,1_s}\ket{\hat n_2,0_s}\ket{1}\ket{0}
+\braket{\hat n_1,1_s}{\Psi_s}\braket{\hat n_2,1_s}{\hat n_1,1_s}\ket{\hat n_2,1_s}\ket{1}\ket{1}
\end{align*}
For $n$ stages this generalizes immediately to
\begin{equation}
\label{wavefunction}
\ket{\Psi_{final}}
=
\sum_{\forall \gamma}
\big(\prod\limits_{i=1}^{n} \braket{\hat n_i,\gamma_i}{\hat n_{i-1},\gamma_{i-1}}\big) \ket{\gamma_{s,f},\gamma}
=\sum _{\forall \gamma} A(\gamma)\ket{\gamma_{s,f},\gamma}
\end{equation}
where in $\ket{\gamma_{s,f},\gamma}$,
$\gamma_{s,f}$ denotes the position of the particle at
the end of the path $\gamma$,
which is that corresponding to the last bit in the chain,
and the second $\gamma$ denotes the joint state of the $n$ ancillas,
the first ancilla being in the state corresponding to $\gamma_1$,
the second to $\gamma_2$,
and so on.
We can see that $\ket{\Psi_{final}}$ reflects a
superposition over all the possible paths,
and the amplitude corresponding to each path is the amplitude computed earlier.

\section{Measuring the measure \label{sect5}}
Let $E$ be any given event (any given set of histories of our system).
As we have said, our objective is to find an experimental procedure that
will reveal $\mu(E)$, the quantum measure of this event. Specifically,
we seek to relate $\mu(E)$ to the probability of some directly
observable instrument-event or ``outcome''. To that end, we have
introduced a series of ancillas which in a sense watch the particle and
record the path that it follows. We now look for a unitary
transformation on the ancillas, followed by a final projective
measurement with two or more outcomes, so arranged that the probability
of the first outcome will be proportional to $\mu(E)$ by a known factor
of proportionality.

We will start by explaining how to achieve this in a simple example with
histories of length two, and then we will generalize to histories of any
length.

\subsection{A simple case}
Consider, then, the simple case shown in figure \ref{box}.
This is the case of
histories of length two,
so the number of beams for the
particle, the number of ancillas and the length of the chains are all
two.

There are four possible histories,
and one can easily compute their
amplitudes for initial wave-function,
$\ket{\Psi}=\alpha \ket{0_s} +\beta \ket{1_s}$.
The result is shown in table \ref{amp}, while
table \ref{m} records the resulting measures of the $2^4=16$ events
which can be built with these histories.
\begin{table}[h]
\renewcommand{\arraystretch}{1.1}
\begin{center}
\begin{tabular}{|l|l|}
\hline
$\gamma$ & $A(\gamma)$ \\
\hline
00 & $\alpha/\sqrt{2}$ \\
\hline
01 & $\alpha/\sqrt{2}$ \\
\hline
10 & $\beta/\sqrt{2}$ \\
\hline
11 & $-\beta/\sqrt{2}$ \\
\hline
\end{tabular}
\caption{\label{amp}
Amplitudes of all the possible histories of the length 2 system for initial state $\ket{\Psi}=\alpha \ket{0_s} +\beta \ket{1_s}$.}
\end{center}
\end{table}
\begin{table}[h]
\renewcommand{\arraystretch}{1.1}
\begin{center}
\begin{tabular}{|l|l|}
\hline
$E$& $\mu(E)$ \\
\hline
$\{\varnothing\}$ & $0$ \\
\hline
$\{00\}$ & $|\alpha|^2/2$ \\
\hline
$\{01\}$ & $|\alpha|^2/2$ \\
\hline
$\{10\}$ & $|\beta|^2/2$ \\
\hline
$\{11\}$ & $|\beta|^2/2$ \\
\hline
$\{00,01\}$ & $|\alpha|^2$ \\
\hline
$\{00,10\}$ & $|\alpha+\beta|^2/2$ \\
\hline
$\{00,11\}$ & $|\alpha|^2/2+|\beta|^2/2=1/2$ \\
\hline
$\{01,10\}$ & $|\alpha|^2/2+|\beta|^2/2=1/2$ \\
\hline
$\{01,11\}$ & $|\alpha-\beta|^2/2$ \\
\hline
$\{10,11\}$ & $|\beta|^2$ \\
\hline
$\{00,01,10\}$ & $(|\alpha+\beta|^2+|\alpha|^2)/2$ \\
\hline
$\{00,01,11\}$ & $(|\alpha-\beta|^2+|\alpha|^2)/2$ \\
\hline
$\{00,10,11\}$ & $(|\alpha+\beta|^2+|\beta|^2)/2$ \\
\hline
$\{01,10,11\}$ & $(|\alpha-\beta|^2+|\beta|^2)/2$ \\
\hline
$\{00,01,10,11\}$ & $1$ \\
\hline
\end{tabular}
\caption{\label{m}
The 16 possible events $E$ and their measures $\mu(E)$, for a length 2 system with initial state $\ket{\Psi}=\alpha \ket{0_s} +\beta \ket{1_s}$.}
\end{center}
\end{table}
Once the ancillas have done their work, we will measure them in a
suitably chosen basis and interact no further with the particle.
If $x$ is a possible outcome of our measurement and $\ket{x}$ is the
associated eigenvector, then
the probability for outcome $x$ is
\begin{equation}
\label{prob}
P(x) = || \, \mathbb{I}_s\otimes \ket{x}\braket{x}{\Psi} \, ||^2
\end{equation}
For computing $P(x)$ we need the ``final wavefunction'' found above, namely
\begin{equation}
\label{f2}
\ket{\Psi}=\frac{1}{\sqrt{2}}(\alpha\ket{0_s00}+\alpha\ket{1_s01}+\beta\ket{0_s10}-\beta\ket{1_s11})
\end{equation}

\subsubsection{Trivial measures and easy to measure measures}
Among the events that we have shown in table \ref{m} there are two
that are trivial and need not be measured at all: the empty set and the
set of all histories. Almost as trivial are the singleton events, those which
comprise only one history. For these events, we didn't need the
ancillas at all, but since we have them,
it suffices to measure each ancilla separately and observe which chain
results, because that is equivalent to directly observing which path
the particle has followed.
In fact, equation (\ref{f2}) says precisely that
the probability that these measurements yield the chain $\gamma$ is exactly
the measure of the event containing just the history $\gamma$:
\begin{equation}
P(00)=|\alpha|^2/2 \quad P(01)=|\alpha|^2/2 \quad P(10)=|\beta|^2/2 \quad \quad P(11)=|\beta|^2/2
\end{equation}

\subsubsection{Two-history events}
Turning now to events that contain two histories (the first case of real interest),
let's look first at,
$\{00,01\}, \{10,11\}, \{00,10\}$ and $\{01,11\}$.
All of these events have in common that both histories
agree in one bit and differ in the other.
Hence, we want the ancilla that records the bit where they differ
to ``forget'' that information.
To do this, as explained before, we will
measure that ancilla in the basis ($\ket{+}$, $\ket{-}$).
The other ancilla, we will measure in the basis ($\ket{0}$, $\ket{1}$).
The outcome-probabilities we obtain this way are, respectively,
\begin{equation}
P(0+)=|\alpha|^2/2 \quad P(1+)=|\beta|^2/2 \quad P(+0)=|\alpha+\beta|^2/2 \quad P(+1)=|\alpha-\beta|^2/2
\end{equation}
Thus, we recover the desired measures up to a factor of two that
comes from the fact that we have a probability one half of obtaining
$\ket{+}$.
This lost factor of 2 in probability represents the inefficiency of
extracting information that we didn't really need, and then having to
forget it.

It is instructive to compute also the probabilities
where we get outcome $\ket{-}$:
\begin{equation}
P(0-)=|\alpha|^2/2 \quad P(1-)=|\beta|^2/2 \quad P(-0)=|\alpha-\beta|^2/2 \quad P(-1)=|\alpha+\beta|^2/2
\end{equation}
We can see that where there is no interference between the histories
(the first two probabilities) we have obtained the measure again, but where
there is interference the probability doesn't correspond to the true
measure. For this reason we must take the probabilities with outcome
$\ket{+}$.
More generally, for events that may contain more than two histories, we will always look for a
superposition that won't alter the interference among them.

The remaining two-history events are $\{00,11\}$ and $\{01,10\}$, and
for them, we will need to involve both ancillas nontrivially.
For example, we can perform first a unitary operation on the ancillas with
the effect:
\begin{equation}
U:\ket{a}\ket{b} \rightarrow \ket{a\oplus b}\ket{b}
\end{equation}
where with $\oplus$ we denote Boolean addition,
i.e. $0\oplus 0=1\oplus 1=0$, $0\oplus 1=1\oplus 0=1$.
Under this transformation the wavefunction becomes:
\begin{equation}
\ket{\Psi}=\frac{1}{\sqrt{2}}(\alpha\ket{0_s00}+\alpha\ket{1_s11}+\beta\ket{0_s10}-\beta\ket{1_s01})
\end{equation}
If after doing this, we measure the first ancilla in the
basis ($\ket{0}$, $\ket{1}$) and the second one in the basis ($\ket{+}$, $\ket{-}$),
we obtain
\begin{equation}
P(0+)=\frac{1}{4} \quad P(1+)=\frac{1}{4}
\end{equation}
Comparing with the table, we see that we have obtained correctly the measures, $\mu\{00,11\}$ and $\mu\{01,10\}$,
with the same normalization of $1/2$ as before.
The reason this works is that these events can be characterized by the
parities of their chains, $(\gamma_1,\gamma_2)$,
namely $\gamma_1\oplus\gamma_2=0$ for the first event
and $\gamma_1\oplus\gamma_2=1$ for the second event.
By the transformation $U$, we arrange for the first ancilla to hold this
parity, and by our choice of what to measure, we ``erase'' the now
unwanted information held by the second ancilla, which still would
distinguish between the two histories comprising the event.

\subsubsection{Three-history events\label{3}}%
This type of event is sufficiently close to the general case that it
seems best to stop thinking in terms of the separate ancillas, and ask
instead what measurement we would like to perform in their joint Hilbert
space.


Suppose, for example, that we are are interested in the event $E=\{00,01,10\}$.
We then want to measure in an orthonormal basis containing the superposition,
$\ket{00}+\ket{01}+\ket{10}$.
We will take the following basis:
\begin{equation}
\label{b1}
\ket{1}=\frac{1}{\sqrt{3}}(\ket{00}+\ket{01}+\ket{10})
\end{equation}
\begin{equation}
\ket{2}=\frac{1}{\sqrt{3}}(\ket{00}-\ket{01}+\ket{11})
\end{equation}
\begin{equation}
\ket{3}=\frac{1}{\sqrt{3}}(\ket{00}-\ket{10}-\ket{11})
\end{equation}
\begin{equation}
\label{b4}
\ket{4}=\frac{1}{\sqrt{3}}(\ket{01}-\ket{10}+\ket{11})
\end{equation}
We claim that $\mu(E) = \mu\{00,01,10\}$ is deducible from the
probability of obtaining the measurement-outcome $\ket{1}$.
What is important here is that $\ket{1}$ corresponds to a
superposition of the three histories of the event with the same
weight and with no phase between them. Any phase that was present would affect the
way the different histories interfere, as happened before when we
looked for the vector $\ket{-}$.
The probabilities of the four outcomes are
\begin{align*}
&P(1)=\frac{1}{6}(|\alpha+\beta|^2+|\alpha|^2) \quad P(2)=\frac{1}{6}(|\alpha+\beta|^2+|\alpha|^2)\\
&P(3)=\frac{1}{6}(|\alpha-\beta|^2+|\beta|^2) \quad P(4)=\frac{1}{6}(|\alpha-\beta|^2+|\beta|^2)
\end{align*}
In particular, we see as claimed that the probability of outcome 1 is
one third of the measure of the event $\{00,01,10\}$ we were looking
for. The factor of 3 comes from the normalization factor
$\frac{1}{\sqrt{3}}$, which in turn just reflects the number of
histories comprising $E$.

The probabilities of outcomes 2 and 3 are not proportional to the measures of the sets of
histories superposed in $\ket{2}$ and $\ket{3}$ because of the phases introduced.
Curiously, however, there is no discrepancy in the case of outcome 4. In that case,
also, a phase is introduced but it is a relative phase between the histories ending on
$0$ and the histories ending on $1$. Since histories with different final positions do
not interfere, such a phase doesn't affect the answer.

In order to measure the measure of one of the remaining two three-history events,
we need to measure the ancillas in a basis including the sum of the
ancilla-states corresponding to the event in question, or else in some other
superposition with phases that cannot affect the probability, as happened with
$\ket{4}$.

We remark here that it wasn't really necessary to perform a ``complete
measurement'' on the ancillas in any basis. It would have sufficed to measure
the ancilla observable that took (say) the value $0$ on $\ket{1}$, and the value
$1$ on its orthogonal complement.

\subsection{The general case}
In a more general situation with histories of $n$ steps,
we will have $2^n$ possible histories and the events will be collections of
them. Suppose we wish to measure the measure of an event $E$ containing $k$
histories:
\begin{equation}
E=\{\gamma^1,\gamma^2,\gamma^3,...\gamma^k\}
\end{equation}
We can use the expression (\ref{mu(X)}) derived earlier to find the
value we are after:
\begin{equation}
\mu(E)=\sum _{l,l'=1}^k A(\gamma^l) \bar{A}(\gamma^{l'}) \delta_{\gamma^{l}_n,\gamma^{l'}_n}
\end{equation}
The most direct approach to determining $\mu(E)$ experimentally is, as we have
done before, to look for the superposition of the $k$ chains contained in this
event, that is, to measure the ancillas in any basis containing the following
state:
\begin{equation} \label{superpos}
\ket{E}=\frac{1}{\sqrt{k}} \sum_{i=1}^k \ket{\gamma^i}
\end{equation}
Provided that the measurement is performed on the ancillas without touching the
system itself,
the probability of outcome $E$
is given by (\ref{prob}) with wavefunction (\ref{wavefunction}).
The projector in this case is
\begin{equation}
\Pi_E=\mathbb{I}_s\otimes \ket{E}\bra{E}=\frac{1}{k}\mathbb{I}_s\otimes \sum_{l,l'=1}^k \ket{\gamma^l} \bra{\gamma^{l'}}
\end{equation}
When we apply this projector to our wavefunction we get:
\begin{align*}
\Pi_E \ket{\Psi_{final}}
=& \big(\frac{1}{k}\mathbb{I}_s\otimes \sum_{l,l'=1}^k \ket{\gamma^l} \bra{\gamma^{l'}}\big)
\sum _{\forall \gamma} A(\gamma)\ket{\gamma_{s,f},\gamma}=\\
=& \frac{1}{k}\sum_{l,l'=1}^k A(\gamma^{l'})\ket{\gamma_{s,f}^{l'},\gamma^l}
\end{align*}
The probability of outcome $E$ is the squared norm of this state:
\begin{align*}
P(E)= || \, \Pi_E \ket{\Psi_{final}} \, ||^2
=&\frac{1}{k^2}\sum_{i,i',j,j'=1}^k A(\gamma^{i'})\bar{A}(\gamma^{j'})\braket{\gamma_{s,f}^{j'},\gamma^j}{\gamma_{s,f}^{i'},\gamma^i}=\\
=&\frac{1}{k}\sum_{i',j'=1}^kA(\gamma^{i'})\bar{A}(\gamma^{j'})\delta_{\gamma^{i'}_n,\gamma^{j'}_n}=\frac{\mu(E)}{k}
\end{align*}
where we have used the orthonormality relations
$\braket{\gamma_{s,f}^{j'},\gamma^j}{\gamma_{s,f}^{i'},\gamma^i}=\delta_{\gamma^{i'}_n,\gamma^{j'}_n}\delta_{i,j}$.
As anticipated, $ P(E)$ is the measure $\mu(E)$ of the event in question, divided by the number of histories in the event.

Thus, we have shown in general that in order to measure the measure of an event
$E$, it suffices to determine the probability of the corresponding superposition
$\ket{E}$ (in the ancillas) of the histories comprising $E$. In principle, this
solves the problem completely. In practise, however, it might not be easy to
find an accessible observable in the Hilbert space of the $n$ ancillas that has
$\ket{E}$ as an eigenvector. In the next section we will see some ways to
simplify this task.

It is also worth taking note of the state of the combined system after the
measurement, which is
\begin{equation} \label{pm}
\begin{split}
\frac{\Pi_E \ket{\Psi_{final}}}{|| \, \Pi_E \ket{\Psi_{final}} \, ||}
&=\frac{1}{\sqrt{k\mu(E)}}\sum_{l,l'=1}^k A(\gamma^{l'})\ket{\gamma_{s,f}^{l'},\gamma^l}
\\&=\big( \frac{1}{\sqrt{\mu(E)}} \sum_{l=1}^k A(\gamma^{l})\ket{\gamma_{s,f}^{l}} \big) \ket{E}
\end{split}
\end{equation}
After a measurement of the ancillas which yields the result $E$, they are of
course no longer entangled with the particle, but what's of interest in (\ref{pm}) is the wave
function of the particle that results from such a measurement.
As is easy to recognize,
it is precisely the wave function that one would obtain in the path-integral
formalism by performing a ``conditional'' integral to which not every history
contributes, but only those histories contained in the event $E$.

\section{Measuring in a big Hilbert space \label{sect6}}
As we have seen, in order to measure the measure we need to look for a particular
superposition in a $2^n$-dimensional Hilbert space. This can be hard to implement,
and in this section we will examine some ways of doing it.

\subsection{Simplification with boolean sum}
First of all let's explain how via 2 qubit gates we can reduce the number of
multi-ancilla measurements we have to do.
We will do this by generalizing the device of Boolean sums that we utilized
earlier. We will start by treating the simple case of two-history events and
then generalize to 3 histories and k histories.

\subsubsection{Two histories}
Suppose the event whose measure we want to measure consists of two histories:
$E=\{\gamma^1,\gamma^2\}$. For any given pair of chains, there will be two kind of bits,
bits that are shared by both chains and bits in which they differ. Since the
order in which the bits occur
is not important here, we can write the chains as
\begin{align*}
\gamma^1=(a_1,a_2,a_3,...a_m,\ &b_1, &{b_2,} \ &b_3, &\cdots &b_{n-m}) \\
\gamma^2=(a_1,a_2,a_3,...a_m,\ &b_1\oplus 1, &{b_2\oplus 1,}\ &b_3\oplus 1, &\cdots &b_{n-m}\oplus 1)
\end{align*}
In accordance with (\ref{superpos}), we thus want to design a measurement which
looks for the state,
\begin{equation}
\ket{E}=\frac{1}{\sqrt{2}}\bigotimes_{i=1}^m\ket{a_i} \big( \bigotimes_{j=1}^{n-m}\ket{b_j} +\bigotimes_{j'=1}^{n-m}\ket{b_{j'}\oplus 1} \big)
\end{equation}

The tensor product structure of this state will let us build up our measurement
from simpler pieces. The first set of factors can be measured directly, qubit by
qubit. The second factor cannot, but we will now show that the required measurement
can also be built up from single qubit measurements.

Recall that in the simple case of $n=2$, we introduced a unitary operation, the
Boolean sum, that allowed us to make do with a single qubit measurement. We can
generalize this idea to $l$ qubits in the following way:
\begin{equation}
U:\ket{b_1}\ket{b_2}...\ket{b_{l-1}}\ket{b_{l}} \rightarrow \ket{b_1 \oplus b_2}\ket{b_2\oplus b_3}...\ket{b_{l-1}\oplus b_l}\ket{b_{l}}
\end{equation}
This unitary operation can be decomposed in $l-1$ Boolean sums which are two
qubit gates, so it can be easily implemented. We apply this unitary to the $n-m$
qubits corresponding to the parts of the chains that differ and get the following
chains:
\begin{align*}
\gamma^1=(a_1,a_2,a_3,\cdots a_m,&b_1 \oplus b_2&,&b_2\oplus b_3&,&b_3\oplus b_4&,\cdots&b_{n-m})\\
\gamma^2=(a_1,a_2,a_3,\cdots a_m,&b_1 \oplus b_2&,&b_2\oplus b_3&,&b_3\oplus b_4&,\cdots&b_{n-m}\oplus 1)
\end{align*}
Now the chains differ just in one qubit! Therefore we can measure individually
each one of the common qubits in the $(\ket{1},\ket{0})$ basis and measure the
last qubit in the $(\ket{+},\ket{-})$ basis, so as to ``forget it'',
as we have explained before.

Let us prove that measuring this way after performing the unitary operation is
equivalent to measuring for the original superposition $\ket{E}$. The probability
doesn't change when a state evolves unitarily if the projector also evolves
unitarily:
\begin{equation}
P(E) = || \, \Pi_E \ket{\Psi_{final}}||^2 = || \, U\Pi_E U^{-1} U \ket{\Psi_{final}}||^2
\end{equation}
The unitary evolution of our projector is:
\begin{equation}
U\Pi_EU^{-1}=\mathbb{I}_s\otimes U\ket{E}\bra{E}U^{-1}
\end{equation}
where:
\begin{align*}
U\ket{E}=&\frac{1}{\sqrt{2}} \bigotimes_{i=1}^m\ket{a_i} U
\big( \bigotimes_{j=1}^{n-m}\ket{b_j} +\bigotimes_{j'=1}^{n-m}\ket{b_{j'}\oplus 1} \big)=\\
=&\frac{1}{\sqrt{2}}\bigotimes_{i=1}^m\ket{a_i} \bigotimes_{j=1}^{n-m-1}\ket{b_j \oplus b_{j+1}}
\big(\ket{b_{n-m}} + \ket{b_{n-m}\oplus 1} \big)=\\
=&\bigotimes_{i=1}^m\ket{a_i} \bigotimes_{j=1}^{n-m-1}\ket{b_j\oplus b_{j+1}} \ket{+}
\end{align*}
which is exactly the state we were proposing to measure.
Therefore
measuring for this state after the Boolean sum is an equivalent procedure.
But after the
sum, the required measurement is no longer a measurement of some abstract
observable in a big Hilbert space, but $n$ individual measurements of simple
observables of each ancilla.

\subsubsection{Three histories}%
Suppose now that we want to measure the measure of an event consisting of three
histories, $E=\{\gamma^1,\gamma^2,\gamma^3\}$. Now we will have at most 4 kinds of
bits: bits shared by every chain, bits that are different in the first chain,
bits that are different in the second chain and bits that are different in the
third chain:
\begin{align*}
\gamma^1=(a_1,...a_m,\ &b_1&,...&b_{p},\ &c_1&,...&c_{q}, \ &d_1&,...&d_{n-(m+p+q)})\\
\gamma^2=(a_1,...a_m,\ &b_1\oplus 1&,...&b_{p}\oplus1,\ &c_1\oplus1&,...&c_{q}\oplus1,\ &d_1&,...&d_{n-(m+p+q)})\\
\gamma^3=(a_1,...a_m,\ &b_1\oplus 1&,...&b_{p}\oplus1,\ &c_1&,...&c_{q},\ &d_1\oplus1&,...&d_{n-(m+p+q)}\oplus1)\\
\end{align*}
For the subchains $b$, $c$ and $d$ we can do as before, apply the Boolean sum
over each subspace to make each chain differ only in the last bit:
\begin{align*}
U\gamma^1=(a_1\oplus a_2,...a_m,\ &b_1\oplus b_2&,...&b_{p}, \ &c_1\oplus c_2&,...&c_{q}, \ &d_1\oplus d_2&,...&d_{n-(m+p+q)})\\
U\gamma^2=(a_1\oplus a_2,...a_m,\ &b_1\oplus b_2&,...&b_{p}\oplus1, \ &c_1\oplus c_2&,...&c_{q}\oplus1, \ &d_1\oplus d_2&,...&d_{n-(m+p+q)})\\
U\gamma^3=(a_1\oplus a_2,...a_m,\ &b_1\oplus b_2&,...&b_{p}\oplus1, \ &c_1\oplus c_2&,...&c_{q}, \ &d_1\oplus d_2&,...&d_{n-(m+p+q)}\oplus1)\\
\end{align*}
Then, instead of having to measure for the superposition of the three chains in
the bigger Hilbert space, we can break the measurement into $n-3$ individual
measurements plus a measurement in the Hilbert space of three ancillas looking for
the state,
\begin{equation}
\ket{\Psi}
=\frac{1}{\sqrt{3}}
(\ket{b_p, c_q, d_{n-(m+p+q)}}+ \ket{b_p\oplus 1, c_q\oplus 1, d_{n-(m+p+q)}}+\ket{b_p\oplus 1, c_q, d_{n-(m+p+q)}\oplus 1})
\end{equation}

\subsubsection{$k$ histories}
These results generalize as follows to events of $k$ histories. We are always
able to cut the chains in a similar fashion as we have done above and apply a
Boolean sum that makes each set of subchains differ just in one bit. By doing
so, we reduce the single measurement in the $2^n$-dimensional Hilbert space to
$n-\alpha$ measurements on individual qubits and a single measurement in an
$2^\alpha$-dimensional Hilbert space, where $\alpha$ is the number of different
subchains.

For each event-cardinality $k$, we can bound the possible values of $\alpha$,
both above and below.
As a perusal of the array beginning the previous subsection will reveal,
a lower bound on $\alpha$ is the number of bits necessary to distinguish $k$
histories. With $m$ bits we can label $2^m$ different histories, so for
labelling $k$ histories we will need at least the base-2 logarithm of $k$ bits:
\begin{equation}
\alpha \geq \lceil \log_2 k \rceil
\end{equation}
where $\lceil . \rceil$ denotes the ceiling function, i.e. the function that
rounds its argument up to the next bigger or equal integer. For example for
$k=3$, the logarithm is between 1 and 2, so we will need at least 2 bits.

To derive an upper bound is a bit more complicated. If we take as a reference
one particular
chain we can start counting how many differing subchains we can make. We will
have $k-1$ subchains for which one of the other chains is different but the
others are still like the first one. There are $\binom{k-1}{2}$ subchains for
which two chains differ from the first one while the rest are same... In general
there will be $\binom{k-1}{i}$ subchains for which $i$ chains differ from the
first one while the rest are same. For counting the total number of subchains we
have to add over all this possibilities:
\begin{equation}
\sum_{i=1}^{k-1} \binom{k-1}{i}=\sum_{i=0}^{k-1} \binom{k-1}{i}-1=2^{k-1}-1
\end{equation}
We have also to take into account that we have $n$ ancillas, which is a bound
that has to be satisfied. In sum, $\alpha$ must lie in the range,
\begin{equation}
\lceil \log_2 k \rceil \leq \alpha \leq \min{(2^{k-1}-1,n)}
\end{equation}
where $\alpha$ is the number of ancillas we have to measure together in a
superposition state after doing the Boolean sums. Notice that it can grow
exponentially with $k$ until it reaches its bound $n$. For such cases, and for
cases with $k > 2^{n-1}$ we cannot break our measurement into simpler ones by
applying Boolean sums, and we are thrown back to measuring a superposition in the
whole Hilbert space.

\subsection{Measuring a superposition}
\begin{figure}
\centering
\includegraphics[scale=0.65]{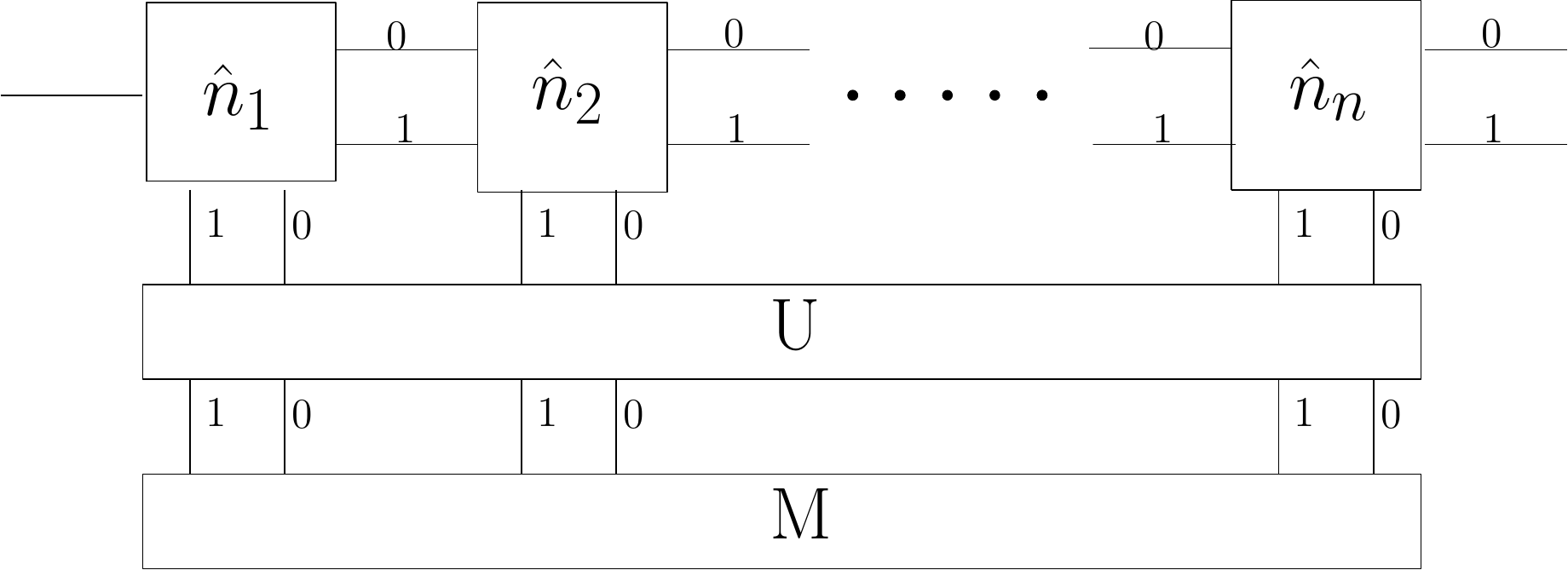}
\caption{\label{meas}Sketch of how one might realize a given measurement for $n$
ancillas. First we implement a unitary gate over the $n$ ancillas and then we
subject them to a more accessible measurement. The latter can consist of
individual measurements on each ancilla or the measurement of a global property
pertaining to the ancillas jointly.}
\end{figure}

Let us approach the question from a somewhat different angle. We want to measure
a state $\ket{E}$ which is a superposition of product states. If we measured
each ancilla individually in the basis ($\ket{0}$,$\ket{1}$), we wouldn't be
able to access this superposition. Instead, as indicated in figure \ref{meas},
we can try to invent a unitary transformation in the space of the ancillas which
will map the state $E$ into something which we can measure easily, like a
particular chain of bits or an eigenstate of some global operator.

\subsubsection{Transforming $E$ into a single chain \label{transforming}}%
\begin{figure}
\centering
\includegraphics[scale=1]{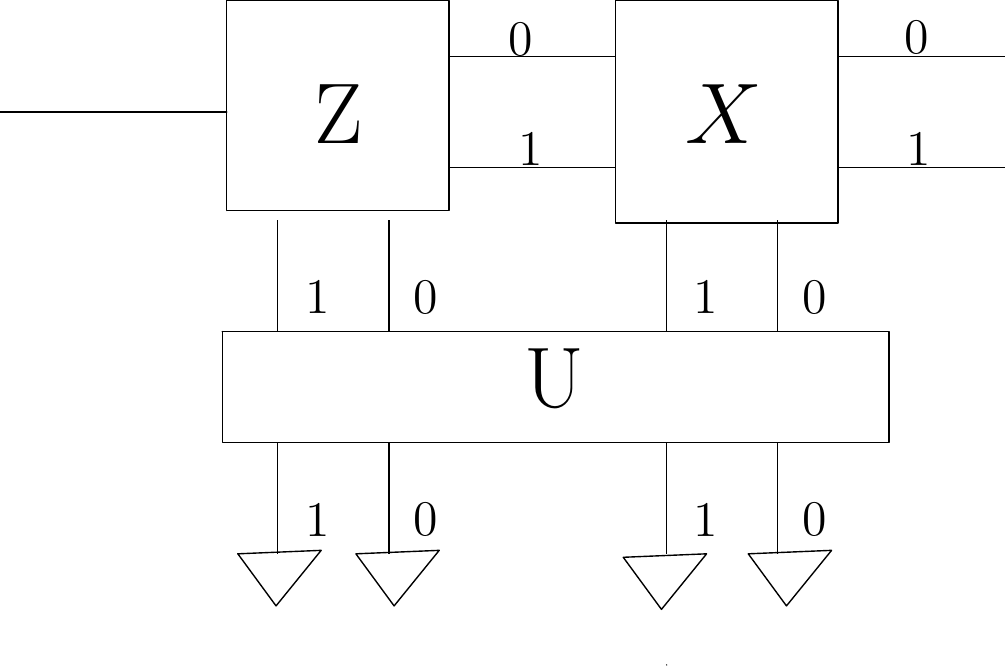}
\caption{\label{2ancilla} A setup with histories of length 2. We couple ancillas to both
the $Z$ and $X$ devices, apply a unitary transformation to the two ancillas, and then
measure each ancilla in the basis ($\ket0$,$\ket1$).}
\end{figure}

Any unitary operator that mapped the state $\ket{E}$ (or the resulting state
after doing some Boolean sums as above) into a product state of the ancillas
would simplify life. But how to design and implement such a unitary? Consider
once again, for example, the measurement proposed in $\S$ \ref{3} for the
three-history events in the case of a length 2 system. A way to perform a
measurement in the basis of equations (\ref{b1}-\ref{b4}) is to implement a
unitary operation that takes the state $\ket{1}$ to the state $\ket{00}$, the
state $\ket{2}$ to the state $\ket{01}$, and so on. We show this schematically in
figure~\ref{2ancilla}.

For each event $E$ we want to measure there are infinitely many bases that
contain the state $\ket{E}$, so there are infinitely many unitary operations that
would allow us to measure the measure (a unitary transformation being equivalent
to a change of basis). An interesting unitary of this sort is a quantum fourier
transform in the subspace of the $k$ histories that form our event. Our event is
\begin{equation}
E=\{\gamma^1,\gamma^2,\gamma^3,...\gamma^k\}
\end{equation}
We define the Fourier transform as
\begin{align}
&U:\ket{\gamma^j}\rightarrow \frac{1}{\sqrt{k}}\sum_{l=1}^k e^{\frac{2\pi i}{k}jl}\ket{\gamma^l} \forall \gamma^j \in E \\
&U:\ket{\gamma '}\rightarrow \ket{\gamma '} \forall \gamma ' \notin E
\end{align}
(This transformation only needs to act on the qubits that are different after any Boolean
sum we might have done.)
We can check that it is unitary:
\begin{align}
U U^{\dagger}\ket{\gamma^j}
&=\frac{1}{\sqrt{k}}\sum_{l=1}^k e^{\frac{2\pi i}{k}jl}U^{\dagger}\ket{\gamma^l}= \frac{1}{\sqrt{k}}\sum_{l=1}^k
e^{\frac{2\pi i}{k}jl}\big(\frac{1}{\sqrt{k}} \sum_{l'=1}^k e^{-\frac{2\pi i}{k}l'l}\ket{\gamma^{l'}} \big)=\\
&=\frac{1}{k}\sum_{l,l'=1}^k e^{\frac{2\pi i}{k}(j-l')l}\ket{\gamma^{l'}}=\sum_{l'=1}^k \delta_{jl'}\ket{\gamma^{l'}}=\ket{\gamma^j}
\end{align}
where we have used that $\sum_{l=1}^k e^{\frac{2\pi i}{k}al}=k\delta_{a,mk}$, for
$m\in \Z$. Since $U$ is the identity for the subspace of histories not contained in
$E$, we conclude that our Fourier transform is unitary.

Since we are looking for the state $\ket{E}$ we need to know how it transforms under $U$:
\begin{align}
U\ket{E}&=\frac{1}{\sqrt{k}}\sum_{l=1}^k U\ket{\gamma^l}=\frac{1}{k}\sum_{l,l'=1}^k e^{\frac{2\pi i}{k}ll'}\ket{\gamma^{l'}}=\\
&=\frac{1}{k}\sum_{l'=1}^k k\delta_{l',k}\ket{\gamma^{l'}}=\ket{\gamma^k}
\end{align}
Therefore,
measuring for the state $E$,
is equivalent to
measuring for the history $\gamma^k$
after applying the Fourier transform.
Since the latter can be done simply by
measuring each ancilla individually in its ($\ket0$,$\ket1$) basis, we have
found a way to measure the measure of $E$ with individual
ancilla measurements.
%
The difficulty is now in performing the unitary
transformation which acts as a Fourier transform, but only
in the subspace of histories of the event $E$.

\subsubsection{Global Observables}
Instead of trying to use a fourier transform to reduce the measurement of
$\ket{E}\bra{E}$ to something more manageable, we could think to measure directly
a suitable global observable in the bigger Hilbert space. If with a
unitary transformation we could map the state $\ket{E}$ to a particular
eigenstate of the global observable, we could look for $\ket{E}$
just by measuring the global observable.

For example, if our ancillas were themselves spin 1/2 particles, instead of
measuring each of their spins in the $z$ direction, we could measure their total
spin (or the total spin of just a few of them), together with its projection in
the $z$ direction. In a three-history event, for illustration, after applying
the Boolean sums described earlier, we need to measure a superposition in two
qubits. Let us map $\ket{E}$ to the singlet state and then measure the total
spin. The probability of obtaining spin 0 as a result will then give us
the measure of $E$ (when corrected with the adequate factor).

Generalizing this idea to arbitrarily many histories
seems to be highly nontrivial. In the best case, we might find an observable
for which one eigenvalue was a singlet while the other was a multiplet with
degeneracy $2^n-1$. In that case, with an adequate unitary operation, we could
describe our measurement with the following projectors:
\begin{equation}
\label{proj1}
\Pi_E=\ket{E}\bra{E}
\end{equation}
\begin{equation}
\label{proj2}
\Pi_{\bar{E}}=\mathbb{I}-\ket{E}\bra{E}
\end{equation}

\section{Interpretation \label{sect7}}
We have designed an experimental setup that will allow us to ``measure the measure''
of any event $E$. More specifically, we have identified the measure
$\mu(E)$ with the probability of a certain experimental outcome $O(E)$ corrected
by a known factor.
Now a question we can ask ourselves is if obtaining the outcome $O(E)$ means that
the event $E$ has really happened, and conversely if not obtaining it means that
$E$ did not happen.

Here we need to be careful, since most formulations of quantum theory do not
let one draw conclusions about what has or has not happened microscopically. In
the context of quantum measure theory, however, it is natural to postulate
that no event of measure 0 can happen. Moreover in certain extensions of this
{\it\/preclusion postulate\/} (including the so called multiplicative scheme) it is
sometimes possible to conclude that the complement of a precluded event does happen.


Let us now analyse our whole system (particles and ancillas) from this point
of view, and ask which particle events are compatible with a particular outcome of
our measurements on the ancillas. To start with, let us ask which events have a
measure different from zero, given a particular outcome of our experimental
procedure.

\begin{figure}
\centering
\includegraphics[scale=1]{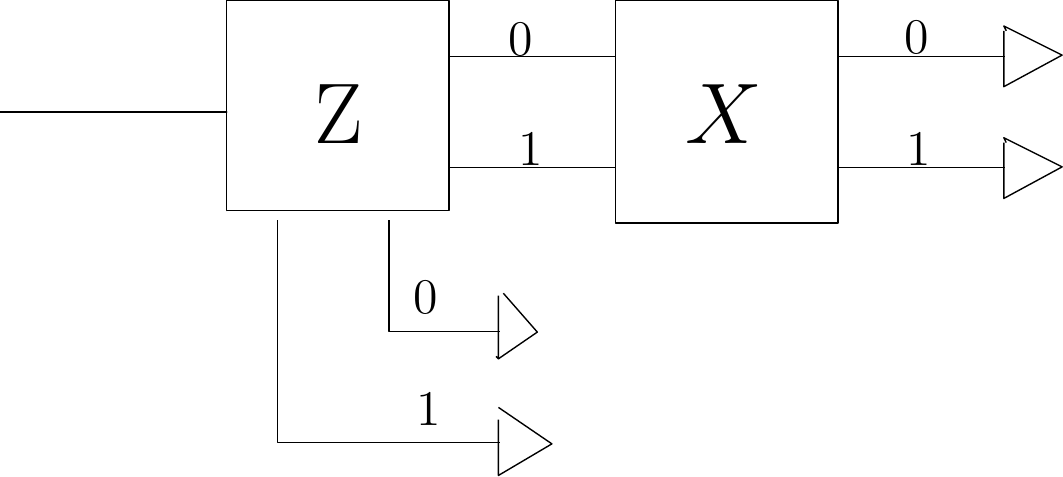}
\caption{\label{1ancilla}
A setup for histories of length 2. We couple an ancilla to the
first analyzer (or its output beams), and then measure it in the
$Z$-basis $(\ket{1},\ket{0})$.
No ancilla is needed for the second analyzer, since it is equivalent to a
strong measurement on the emerging particle.}
\end{figure}

To begin with,
we will analyse the simple arrangement in which the particle passes through
only two analyzers and we have only a single ancilla, as shown in figure
\ref{1ancilla}. Notice that we don't couple an ancilla to the second analyzer
because we can directly observe which beam the particle emerges in. Interposing
an ancilla
would accomplish nothing beyond complicating the notation.
For this setup, the possible histories and their amplitudes
are shown in table \ref{amp2}.

\begin{table}[h]
\renewcommand{\arraystretch}{1.1}
\begin{center}
\begin{tabular}{|l|l|}
\hline
$\gamma _s, \gamma _a$ & $A(\gamma _s, \gamma _a)$ \\
\hline
00,0 & $\alpha/\sqrt{2}$ \\
\hline
01,0 & $\alpha/\sqrt{2}$ \\
\hline
10,0 & $0$ \\
\hline
11,0 & $0$ \\
\hline
00,1 & $0$ \\
\hline
01,1 & $0$ \\
\hline
10,1 & $\beta/\sqrt{2}$ \\
\hline
11,1 & $-\beta/\sqrt{2}$ \\
\hline
\end{tabular}
\caption{\label{amp2}
Amplitudes,
for an initial wave-function $\ket{\Psi}=\alpha \ket{0_s} +\beta \ket{1_s}$,
of all possible histories of a length 2 system
with an ancilla coupled in.
The notation we use for the histories is: we
write first the history of the particle, and then the history of the ancilla, separated
by a comma.}
\end{center}
\end{table}
For the full system of ancilla plus particle, there are eight joint histories in
all. As we might have expected, the table shows that {\it\/every one of them\/}
where the ancilla's history disagrees with the particle's history is precluded.
In this sense, we can affirm that if we measure the ancilla in the
$(\ket{1},\ket{0})$ basis (and also observe the final emerging beam with a
particle-detector), then the particle has actually travelled the path associated
with the outcomes measured.\footnote%
{If we wish to be more cautious, we can say only that the event that the
particle travelled some {\it\/other\/} path than what our measurement
indicated did {\it\/not\/} happen. For example if we measured $(0,0)$
then this, complementary event would comprise the last 7 histories in
the table \ref{amp2}.}

We can generalize this straightforwardly to any number of ancillas. When we
measure the ancillas in their $(\ket{1},\ket{0})$ bases and obtain the outcome $\gamma$,
it is hard to doubt that the particle actually has followed the corresponding path.
That we can deduce the path this way is not surprising, as the setup is equivalent to
doing a strong measurement at each step.

\begin{figure}
\centering
\includegraphics[scale=1]{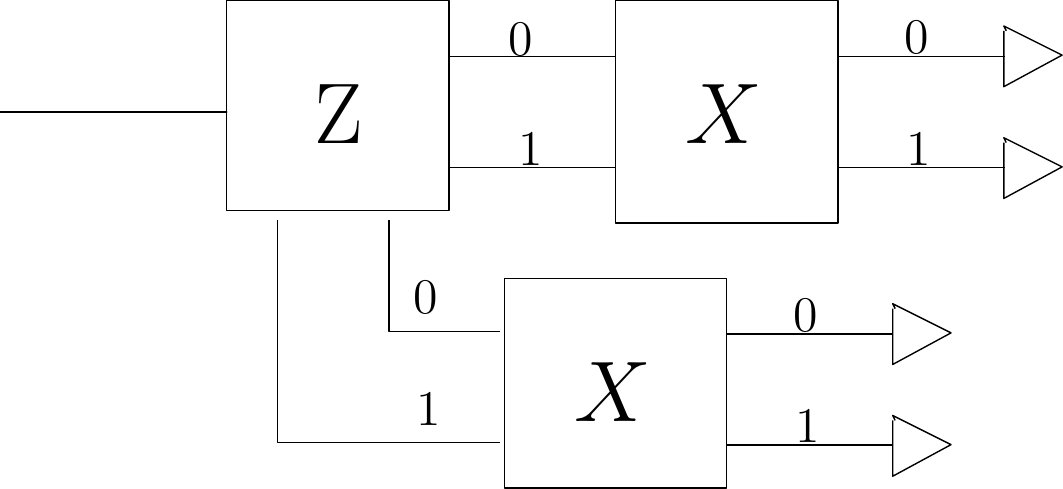}
\caption{\label{1ancilla2}
Another setup for histories of length 2. We couple an ancilla to the
first analyzer (or its output beams), and then measure it in the
$X$-basis, $(\ket{+},\ket{-})$.
No ancilla is needed for the second analyzer, since it is equivalent to a
strong measurement on the emerging particle.}
\end{figure}

Consider now the slightly more complicated case of figure \ref{1ancilla2}.
As we saw earlier,
this setup lets us measure the measures of the events,
$\{00;10\}$ and $\{01;11\}$. (Now we use a semicolon ``;'' to separate
different histories, since we are using the comma ``,'' to separate the history
of the particle from the history of the ancilla).
Here the ancilla's history has also length 2,
and the set of joint histories with amplitudes different from 0 is: \\
$\{(00,00);(00,01);(01,00);(01,01);(10,10);(10,11);(11,10);(11,11)\}$.
Suppose we
want to measure the measure of the particle-event $\{00;10\}$. As we
saw earlier, this corresponds to measuring the the first ancilla in the $\ket{+}$
state and the second ancilla (if we had included it) in the $\ket{0}$
state. (In fact there is no second ancilla, since we have again simplified our protocol
by measuring the final particle location directly.)
This outcome is only compatible with the histories,
$\{00,00\}$ and $\{10,10\}$.
Therefore when we get $+0$,
we can say that the particle-event $\{00;10\}$ has happened.

What happens, though, when we get the outcome $-0$?
One can check that the histories
compatible with this outcome
are $\{00,01\}$ and $\{10,11\}$,
so if we just care about the history of
the particle, we get that the same histories, $\{00\}$ and $\{10\}$, are compatible!
Then we
can say simply that if we measure $0$ in the second qubit, the event $\{00;10\}$ has
happened. Whether we obtain $+$ or $-$ doesn't affect which histories are
compatible, but it does affect the state,
and we don't correctly recover the measure of the event $\{00;10\}$ when the outcome is $-$.

We can extend this analysis to every two-history event where we apply the
unitary trick described before.
After measuring all the ancillas which are meant to be measured in
the ($\ket{0}$,$\ket{1}$) basis, there will be only two histories that are compatible
with whatever outcomes we have obtained.
When we measure the remaining qubit in the
basis ($\ket{+}$,$\ket{-}$),
either of the two possible histories (the one labelled
with 0 and the one labelled with 1) could have happened, so we can say that the event $E$
has happened.
Again, $E$ happens whether we get $+$ or $-$, but we
can only learn its measure from the measured probabilities if we get $+$.

Before we turn to some more complicated setups, we need to agree on a
linguistic convention. In order to explain what we mean, let us first of
all include in each overall history the results of all the final
measurements made upon the ancillas. In the setup just discussed, for
example, we would include either a `$+$' or a `$-$' depending on which
outcome was obtained. Then it may happen, given a set $O$ of measurement
outcomes, that every (overall) history $\Gamma$ in which $O$ happens,
and whose measure is nonzero, has the further property that a certain
particle event $E$ also happens.\footnote%
{We can also express this by saying that every history which
is inside of $O$ but outside of $E$ has measure zero.}
In this case we will allow ourselves
to say that $E$ also happened,
(and that the complementary event ``not $E$'' did not.)
It is this convention that led us to say,
in the first setup that the particle followed the trajectory $\gamma$,
and in the second setup that $E$ happened when either $+$ or $-$
was obtained. We qualify it as a convention because, as self-evident
as it might sound at first hearing, there are reasons
why one might want to replace it with something different. For more on
this point, see the discussion in \cite{Howinter}.

\subsection{events with $k$ histories}
To try to say ``what has happened'' when the event $E$ of interest is formed
by $k>2$ histories is more complicated, since the analysis in that case depends on
which unitary transformations and final measurements we employ.

Recall first the procedure proposed in $\S$\ref{transforming} for the
three-history events of a length-two system. This was a way to
perform a measurement in the basis of equations
(\ref{b1}-\ref{b4})
by means of
a unitary operation that maps the
ancilla state $\ket{1}$ to $\ket{00}$, $\ket{2}$ to
$\ket{01}$, and so on, as was shown schematically in figure~\ref{2ancilla}.
Following an analysis of the histories similar to what we have done before,
it is not hard to check that
the histories compatible with each of the outcomes $00$, $01$, $10$, $11$,
are those contained respectively in the states, $\ket{1}$, $\ket{2}$, $\ket{3}$, $\ket{4}$.
Hence we can say that if our ancilla measurement yields $00$,
then
the particle event $E=\{00,01,10\}$ has happened,
because
no history of nonzero measure
combines the particle-history $11$ with the ancilla outcome $00$.
(For example the history $\{11,1100\}$ has measure 0, where the last two bits represent
the outcome of the ancilla measurement.)
We might have expected this, because
the state $\ket{1}$ was the one we used to measure the measure of $E$.
However,
when the outcome of our ancilla measurement
is something other than $00$ we cannot state that the event $\{00,01,10\}$ has
not happened! For example, the measure of the event containing the particle
histories $00$ and $01$, and the ancilla histories compatible with measuring $\ket{2}$,
is different from $0$ as there is overlap between $\ket{2}$ and the single-history states
corresponding to $00$ and $01$.

Generalizing these conclusions,
we can say that measuring for --- and obtaining --- $\ket{E}$ implies
that the event $E$ has happened, because only the histories contained in $E$ have
measure different from $0$ when we limit ourselves to ancilla histories that have
the outcome $E$. On the other hand, if the outcome state is not
$\ket{E}$ but has a non-zero overlap with the subspace generated by the
single-history states of the histories contained in $E$
we cannot exclude that $E$ or some subevent of $E$ happened.

Now consider the same system and 3-history event $E$, but with a different setup,
such that
we measure the ancillas in the basis given by the quantum Fourier transform. In
this case we can still imagine the setup as the circuit represented in figure \ref{2ancilla},
but with a different unitary operator, acting now
in the subspace spanned by
$\ket{01}=\ket{\gamma^1}$, $\ket{10}=\ket{\gamma^2}$, and $\ket{00}=\ket{\gamma^3}$.
This procedure
corresponds to measuring in the basis,
\begin{equation}
U^{\dagger}\ket{00}=\frac{1}{\sqrt{3}}\big( \ket{00}+\ket{01}+\ket{10}\big)=\ket{E}
\end{equation}
\begin{equation}
U^{\dagger}\ket{01}=\frac{1}{\sqrt{3}}\big( \ket{00}+e^{\frac{2\pi i}{3}}\ket{01}+e^{\frac{4\pi i}{3}}\ket{10}\big)
\end{equation}
\begin{equation}
U^{\dagger}\ket{10}=\frac{1}{\sqrt{3}}\big( \ket{00}+e^{\frac{4\pi i}{3}}\ket{01}+e^{\frac{2\pi i}{3}}\ket{10}\big)
\end{equation}
\begin{equation}
U^{\dagger}\ket{11}=\ket{11}
\end{equation}
Examining the makeup of these vectors, we see that the outcomes $00$, $01$, $10$
are incompatible with the particle travelling the path $11$, while vice versa, the
outcome $11$ is incompatible with any of the particle-paths $00$, $01$, $10$.
Thus, the correlation is fuller in this setup. If the outcome is one of
$00$, $01$, $10$, then we can say that the event $E=\{00,01,10\}$ has happened,
while the history $11$ has not happened. And if the outcome is $11$ then we can
say that the history $11$ has happened, while the event $E$ has not.
However, as before, it is only the outcome $00$ that lets us
recover the measure of the event $E$, and that lets us assert that the
particle's ``collapsed'' wave function is the same as if it had evolved freely but
following only the trajectories contained in $E$.

The same conclusions evidently hold for events with more than three
histories. When (via a suitable unitary) we measure in a basis
containing the ancilla state $\ket{E}$ and we obtain it, we can say that
the event $E$ has happened. When we obtain something different, what we
can say
depends on the basis in which we have measured.
For a general basis we won't be able to tell whether the event has
happened or not, as was the case for the three-history events and the
basis $\ket1$, $\ket2$ $,\ket3,$ $\ket4$.
But for the Fourier transform basis,
whenever (after applying the unitary transformation) we obtain one of the
histories contained in the event, we can say that the event has happened, and if we
obtain a different history then we can say that the event hasn't happened.

In order to see this, we can observe that the states associated with
each outcome are either a superposition of all the histories in the
event when the outcome is such a history itself, or else the same
history as the outcome when it is not in the event:
\begin{align}
&U^{\dagger}\ket{\gamma^j}=\frac{1}{\sqrt{k}}\sum_{l=1}^k e^{\frac{-2\pi i}{k}jl}\ket{\gamma^l} \forall \gamma^j \in E \\
&U^{\dagger}\ket{\gamma'}= \ket{\gamma '} \forall \gamma ' \notin E
\end{align}
In the language of measures, the only compatible histories whose
measures differ from zero when we get an outcome that corresponds to a
history of the event are precisely those contained in the event. On the
other hand, when we obtain an outcome that is not part of the event we
can say that that history has happened, and therefore the histories in
$E$ have not.

As our analysis has demonstrated, when we measure for and obtain the
state $\ket{E}$, the resulting particle wave function is the same as if
the particle had evolved freely, but following only the trajectories in
$E$. One might wonder what wave function results when we obtain an
outcome different from $\ket{E}$. Is there one among these outcomes
such that it is as if the particle had evolved according to all the
trajectories in the complementary event to $E$? This would correspond
to making the state of the ancillas collapse to
\begin{equation}
\ket{\bar{E}}=\frac{1}{\sqrt{2^n-k}} \sum_{\gamma^i \notin E} \ket{\gamma^i}
\end{equation}
We could arrange for this to be among the possible outcomes by
implementing the quantum fourier transform associated with the histories
not contained in $E$.

In table \ref{Int} we summarize the inferences we have arrived at
so far,
presupposing that
the measurement performed is a projective measurement which completely
collapses the wavefunction of the ancillas, so that they are not
entangled with the particle any more.

\begin{table}[h]
\renewcommand{\arraystretch}{1.1}
\begin{center}
\begin{tabular}{|l|l|l|l|}
\hline
State Measured & $E$ happened? & Probability & Final system $\psi$ \\
\hline
$\ket{E}$ & Yes & $\mu(E)/k$ & Histories in $E$ \\
\hline
$\ket{\Psi} \in \text{span}\{\ket{\gamma^1},...\ket{\gamma^k}\}$ & Yes & Not related to $\mu$ & Histories in $E$ altered \\
$\ket{\Psi} \neq \ket{E} $ & & & \\
\hline
$\ket{\Psi} \notin \text{span}\{\ket{\gamma^1},...\ket{\gamma^k}\}$ & Cannot tell & Not related to $\mu$ & Histories in E and \\
$\ket{\Psi} \not\perp \text{span}\{\ket{\gamma^1},...\ket{\gamma^k}\} $ & & &not in E altered \\
\hline
$\ket{\Psi} \perp \text{span}\{\ket{\gamma^1},...\ket{\gamma^k}\}$ & No & Not related to $\mu$ & Histories not in $E$ \\
$\ket{\Psi} \neq \ket{\bar{E}} $ & & & altered \\
\hline
$\ket{\bar{E}}$ & No & $\mu(\bar{E})/(2^n-k)$ & Histories not in $E$ \\
\hline
\end{tabular}
\caption{\label{Int}
Possible states in which one could find the ancillas when performing a
projective measurement that completely collapses the ancilla
wavefunction. We analyse for each case if we can say that the event
$E=\{\gamma^1,...\gamma^k\}$ happened, if the probability of finding
that state is directly related with the measure of the system event or
its compliment, and if the resulting system wave function is the result
of evolving it via certain histories, with or without introducing extra
phases or weights into the path amplitudes.}
\end{center}
\end{table}

Lastly, we can consider the case where we measure a global
ancilla-observable, for example the total spin when the ancillas are
spin-1/2 particles. As described earlier, we would like this
observable's spectrum to consist of a first eigenvalue with a
degeneracy of 1 and a second eigenvalue with a degeneracy of $2^n-1$.
In that case, applying an appropriate unitary operator will set up a
simple measurement for which the projectors corresponding to each
outcome are those of equations (\ref{proj1}) and (\ref{proj2}). If,
then, we obtain the outcome corresponding to $E$ we can say that the
event has happened, but if not we can not say anything.

We can also imagine a situation in which one can discover an observable with
two outcomes and a nontrivial multiplet associated with each outcome. If
the first multiplet has $k$ states, we can look for a unitary
transformation that maps the $k$ ancilla-histories of our event $E$ to
states of this multiplet, and then measure the observable. Obtaining
the outcome associated with this first multiplet will then mean that the
event $E$ has happened, while measuring the other outcome will mean that
the event hasn't happened (in fact that its complementary event has
happened). This setup would be like a detector for the event $E$, but
we wouldn't be able to recover the measure of $E$ from such a
measurement.
%
Moreover, the final wave function of the particle would still be
entangled with the ancillas, so one could not describe it with the
wavefunction generated by evolving the particle through some specified
subset of the trajectories,
and a density matrix would be a better description.

\section{Conclusions \label{sect8}}
Given any system-event $E_{system}$, we have provided a set of ancillas,
couplings of them to the system and to each other, and an ancilla-event
$E_{ancilla}$ which in a certain sense asks whether $E_{system}$ has happened.
If an ensemble of ``identically prepared" copies of the ``system'' (and
also of the ancillas) is available,
then we can ``measure the measure" of $E_{system}$ by
performing
on the ensemble
projective measurements
that look for $E_{ancilla}$. If $P$ is the
probability of a positive outcome, then $\mu(E_{system}) = k P$, where the
correction factor $k$ is the number of histories comprising
$E_{system}$. Here of course, $\mu(E)$ means the measure of $E$
computed in the absence of ancillas, i.e. for the closed system.
Furthermore, when our ancilla measurement yields a positive outcome, i.e. when
$E_{ancilla}$ happens, then the effective wave function for the system will be
obtained --- if we employ the usual collapse rule --- by propagating its initial
wave-function (or density-matrix) forward via the histories in $E_{system}$.

In light of these results, we can claim in some informal sense that our procedure
constitutes a way to verify that the system-event $E$ has happened without
disturbing the system more than necessary. One might say that we learn that $E$
happened but we learn no more than this.

If we wish to speak more precisely, we can observe that the couplings induce
at the level of the measure a certain correlation between $E_{ancilla}$ and
$E_{system}$, namely the preclusion ($\mu=0$) of the event,
$E_{ancilla}\cap E_{system}^c$, where the superscript $c$ denotes complement.
In words, the event, ``$E_{ancilla}$ but not $E_{system}$'' cannot happen.
If we could employ classical inference then we could conclude that
$E_{ancilla}\implies E_{system}$, however this doesn't necessarily follow
quantum mechanically. Nevertheless we have in our presentation spoken as if a
form of this implication could be assumed.
We also pointed out in this connection, that the ``converse" event,
``$E_{system}$ but not $E_{ancilla}$'',
is {\it\/not\/} precluded in general. Thus we do not
claim, even informally, that the complementary outcome $E_{ancilla}^c$ implies
that $E_{system}$ has {\it\/not\/} happened, or that its complement
$E_{system}^c$ has.

It's worth noting that the procedures we have described are already fairly
realistic. For the kind of system we have discussed, they are not that far from
letting us actually measure the quantum measure of many events $E$ pertaining to
the system. The fact that this is possible, even in principle, lends a direct
experimental meaning to the quantum measure, similarly to how more familiar
schemes of measurement lend experimental meaning to the expectation values of
projection operators. In both cases, one has converted a formally defined
quantity into a macroscopically accessible number.

In saying this, though, we don't mean to imply that the quantum measure has no
meaning other than this experimental one. On the contrary, its real purpose is
to let one reason directly about the quantum world in itself, without the aid of
external observers. But knowing, as we now do, that a direct experimental
determination of the measure is also available can only serve to encourage the
larger interpretive project.

Our procedure generalizes the ``quantum eraser'' setup, in which an ancilla is
coupled to a double slit experiment such that the interference appears or disappears
depending on the basis you measure the ancilla in.
In our case, we have generalized that idea to let us select from all the possible
trajectories, just the subset we are interested in, the subset that constitutes
the event $E$.

In reference \cite{Gudder2011b}, the authors studied a two-site quantum random
walk that is equivalent to the particular case of our experiment in which the
analyzers are placed in the sequence $Z,Y,-Z,-Y,Z,Y,-Z,-Y\dots$. Coupling and
ancillas to this random walker and then measuring them in a similar way to the
one proposed here would allow one to verify experimentally all the properties of
the measure of the system described in that article.

The system we studied herein was rather special, but generalizing our procedure
to any other discrete system would present no difficulty, at least if one
idealizes every sort of ancilla coupling and every unitary as physically
realizable. For continuous systems, one can think of coupling ancillas also with
continuous degrees of freedom, but in order to define continuous trajectories one
would need to couple an ancilla at each instant of time, which would require
infinitely many ancillas. Something similar was done in \cite{PhysRevA.36.5543}
by coupling one ancilla to the system in every time interval $\tau$ and then
taking the $\tau\to0$ limit. In that work, the authors were
interested in how continuous measurements would affect the evolution of expectation
values, so they measured the ancillas immediately, without
interpolating any unitary operator in the way presented here.

\section{Acknowledgements}

\begin{acknowledgements}
AMF would
like to thank his PSI partners for their useful discussions and the long
hours working together.
This research was supported in part by NSERC through grant RGPIN-418709-2012.
This research was supported in part by Perimeter Institute for
Theoretical Physics. Research at Perimeter Institute is supported
by the Government of Canada through Industry Canada and by the
Province of Ontario through the Ministry of Economic Development
and Innovation.

\end{acknowledgements}

\bibliographystyle{spmpsci} 
\bibliography{references} 


\end{document}

%% file: PSIpackages.tex
\usepackage{graphicx}
\usepackage{amsmath}
\usepackage{amsfonts}
\usepackage{amssymb}
\usepackage[sectionbib]{chapterbib}
\usepackage{appendix}
\usepackage{color}
\usepackage{cancel}
\usepackage{tensind}

\usepackage{tensor}
\usepackage{dsfont} 
\usepackage{enumitem}

%% file: PSImacros.tex
\newcommand{\braket}[2] {\langle #1 | #2 \rangle}

\renewcommand\thesection{\arabic{section}}

\newcommand{\be}{\begin{equation}}
\newcommand{\ee}{\end{equation}}
\newcommand{\ba}{\begin{eqnarray}}
\newcommand{\ea}{\end{eqnarray}}

\def \bes {\begin{eqnarray}}
\def \ens {\end{eqnarray}}

\newcommand{\R}{\mathbb R}

\newcommand{\Z}{\mathbb Z}
\newcommand{\C}{\mathbb C}

\newcommand{\beq}[1]{\begin{equation}\label{#1}}
\newcommand{\eeq}{\end{equation}}

\newcommand{\ket}[1]{\left.\left| #1 \right. \right\rangle}
\newcommand{\bra}[1]{\left\langle \left. #1 \right|\right.}
\everymath{\displaystyle}

